\documentclass[10pt,final,journal,letterpaper,twoside,twocolumn]{IEEEtran}
\usepackage{amsfonts}
\usepackage{amssymb}
\usepackage{amsmath}
\usepackage{dsfont}
\usepackage{graphicx,subfigure}
\usepackage{enumerate}
\usepackage{cite}
\usepackage{bigstrut}
\usepackage{float}
\usepackage{balance}
\usepackage{lipsum}
\usepackage{psfrag}
\usepackage{algorithm}
\usepackage[noend]{algpseudocode}
\usepackage[export]{adjustbox}

\newtheorem{lemma}{Lemma}

\begin{document}

\title{Wireless Federated Learning (WFL) for 6G Networks - Part II: The  Compute-then-Transmit NOMA Paradigm}
\author{Pavlos S. Bouzinis, \IEEEmembership{Student Member,~IEEE}, Panagiotis D. Diamantoulakis,  \IEEEmembership{Senior Member,~IEEE}, and George K. Karagiannidis,  \IEEEmembership{Fellow,~IEEE}
\thanks{P. S. Bouzinis, P. Diamantoulakis,  and G.K.  Karagiannidis are with Wireless Communication and Information Processing  Group (WCIP), Department of Electrical and Computer Engineering, Aristotle University of Thessaloniki, Greece, E-mails: \{mpouzinis, padiaman, geokarag\}@auth.gr}}
\maketitle

\begin{abstract}
As it has been discussed in the first part of this work,  the utilization of advanced multiple access protocols and the joint optimization of the communication and computing resources can facilitate the reduction of delay for wireless federated learning (WFL), which is of paramount importance for the efficient integration of WFL in the sixth generation of wireless networks (6G). To this end, in this second part we introduce and optimize a novel communication protocol for WFL networks, that is based on non-orthogonal multiple access (NOMA). More specifically, the \textit{Compute-then-Transmit NOMA (CT-NOMA)} protocol is introduced, where users terminate concurrently the local model training and then simultaneously transmit the trained parameters to the central server. Moreover, two different detection schemes for the mitigation of inter-user interference in NOMA are considered and evaluated, which correspond to fixed and variable decoding order during the successive interference cancellation process. Furthermore, the computation and communication resources are jointly optimized for both considered schemes, with the aim to minimize the total delay during a WFL communication round. Finally, the simulation results verify the effectiveness of CT-NOMA in terms of delay reduction, compared to the considered benchmark that is based on time-division multiple access.
\end{abstract}
\vspace{-0.1in}
\section{Introduction}
\IEEEPARstart{I}{n} Part I of this two parts paper, we presented the advantages, the reference architecture and three core applications of wireless federated learning (WFL), the main challenges toward its efficient integration in the sixth generation of wireless networks (6G), and the identified future directions \cite{part1}. This second  part is motivated by the fact that, as it has  been thoroughly discussed in II. A and III.A of the first part, the utilization of advanced multiple access protocols and the joint optimization of the communication and computing resources can facilitate WFL to meet the stringent latency requirements of 6G networks \cite{part1, konevcny2016, 6G}. To this direction, the use of non-orthogonal multiple access (NOMA) has been proposed in the first part, due to offering low-latency and improved fairness by serving multiple users in the same resource block, in opposition to  orthogonal multiple access (OMA), where each user occupies a single resource block \cite{ding2017}. 
\par  The aim of the second part of this paper is to investigate the potential of using NOMA in WFL, by focusing on a specific paradigm. In more detail, we introduce the \textit{Compute-then-Transmit NOMA (CT-NOMA)} protocol, which is based on the use of two phases during a WFL round. Specifically, users firstly execute and complete the local model training concurrently and  afterwards upload the trained model to the server simultaneously via NOMA. Moreover, we consider two detection schemes for the implementation of  successive interference cancellation (SIC) process in NOMA that correspond to the utilization of a fixed or variable decoding order. The later, is implemented through the time-sharing (TS) strategy, which improves the performance of uplink NOMA by achieving any point of the multiple access channel (MAC) capacity region \cite{goldsmith}. Furthermore, we minimize the total delay of a communication round subject to users' energy constraints for both considered schemes, by jointly optimizing the available computation and communication resources, i.e., users' CPU frequency for local training, the transmission energy and the time intervals for local computations and parameter transmission.  Finally, simulation results demonstrate the effectiveness of the proposed protocol in reducing the delay of a WFL communication round, compared the considered benchmark that is based on time-division multiple access (TDMA).
\vspace{-0.1in}
\section{System Model and Proposed Protocols} 
\subsection{Wireless federated learning model}
Based on the reference architecture that has been presented in detail in the first part of this work, we consider a WFL learning system which consists of $N$ users indexed as $n \in \mathcal{N}=\{1,2,...,N\}$ and a server/base station (BS). Each user $n$ has a local dataset $\mathcal{D}_n$,  where $D_n=\vert \mathcal{D}_n \vert$ are the total data samples. Next we briefly describe the steps of an arbitrary, $i$-th communication round, between the users and the BS in the WFL system.
\begin{enumerate}[i)]
	\item The BS broadcasts the global parameter $\boldsymbol{w}^i$ to all users during the considered round.
	\item After receiving the global model parameter, each user $n \in \mathcal{N}$, trains the local model through its dataset, and then uploads the trained local parameter $\boldsymbol{w}^{i+1}_n$ to the server.
	\item After receiving all the local parameters, the server aggregates them, in order to update the global model parameter  $\boldsymbol{w}^{i+1}$.
\end{enumerate}
The whole procedure is repeated until the global model converges. Moreover, during the first round the server initializes  $\boldsymbol{w}^0$.
\subsection{Compute-then-Transmit NOMA}
\begin{figure}[h!]
\vspace{-0.1in}
	\centering
	\includegraphics[width=0.8\linewidth]{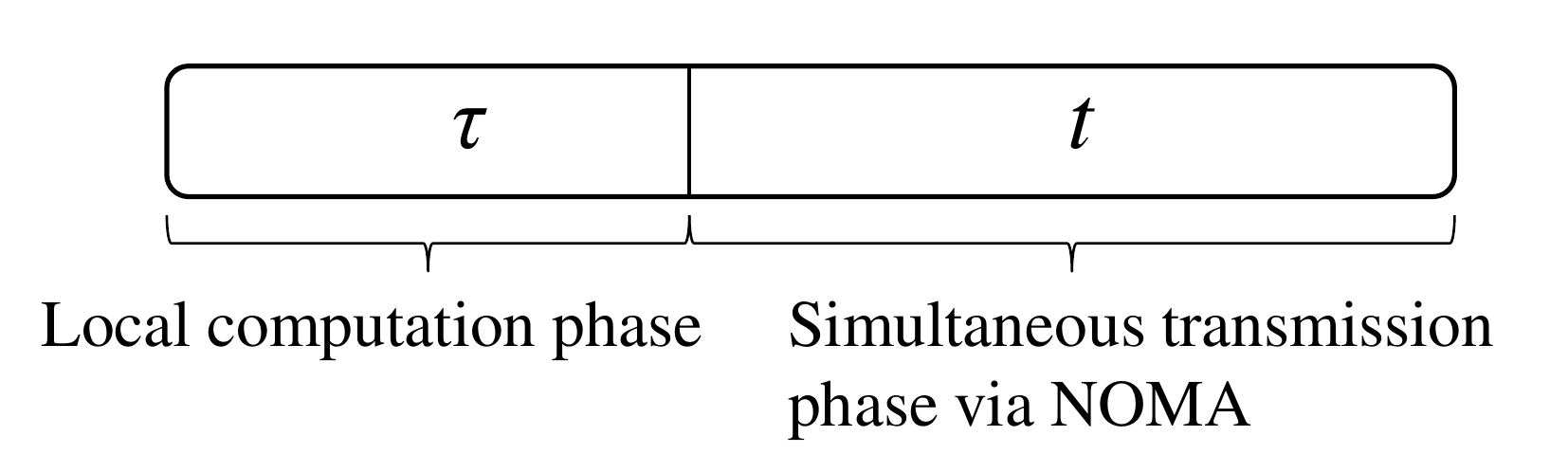}
	\vspace{-0.1in}
	\caption{\textit{Compute-then-Transmit NOMA} during a WFL round.}
	\label{protocol}
	\vspace{-0.2in}
	\label{CTNOMA}
\end{figure}

As it has already been mentioned, CT-NOMA is based on the use of two consecutive phases during a WFL round, i.e., the computation and the communication phase. Specifically, in the first phase users  execute the local computations, while in the second phase they transmit their messages, i.e., the trained parameters, to the BS. During the information transmission phase, the NOMA protocol is considered. According to NOMA, users are capable of transmitting their messages simultaneously, while the BS decodes the users' messages by utilizing SIC. As implied by the term CT-NOMA, all users begin the information transmission procedure via NOMA at the same time instant. Therefore, all users are constrained to  complete the local computations before the information transmission phase, i.e., within time duration  $\tau$, while the transmission phase duration is denoted as $t$. The CT-NOMA protocol is illustrated in Fig. \ref{CTNOMA}.
\par
The utilized computation resources for local model training,
i.e., CPU cycle frequency, from the $n$-th user is denoted as $f_n$. The number of CPU cycles for the $n$-th user to perform
one sample of data in local model training is denoted by $c_n$. Hence, the computation time dedicated for a local iteration is given as
\begin{equation}
\tau_n=\frac{c_nD_n}{f_n}, \quad \forall n \in \mathcal{N}.
\end{equation}
 Accordingly, the energy consumption for a local iteration, can be expressed as follows
\begin{equation}
E^{\mathrm{comp}}_n=\zeta c_n D_nf^2_n = \zeta \frac{c^3_nD^3_n}{\tau_n^2}, \quad \forall n \in \mathcal{N},
\end{equation}
where $\zeta$  is a constant parameter related to the hardware
architecture of device $n$. 
As it was discussed previously, all users are enforced to complete the local computations within $\tau$, with the corresponding energy that is consumed by each user being a decreasing function with respect to $\tau$.
Thus, it should hold
\begin{equation}
	\tau_n=\tau, \quad \forall n \in \mathcal{N}.
\end{equation}

\par
 In the continue, the two proposed schemes regarding information transmission will be described, termed as NOMA with fixed decoding order (FDO)  and NOMA with TS.
\subsubsection{NOMA with FDO} Fixed decoding order refers to the classical uplink NOMA protocol with SIC, which is widely used in the literature \cite{ding2017}. Let $\pi_n$ be the the $n$-th user's decoding order position at the BS, while without loss of generality we consider that $\pi_1<\pi_2<...<\pi_N$. According to NOMA, for decoding the first user's message, interference is created by the residual users, i.e., $n=2,...,N$, while on the second user's message, interference is created by the users indexed as $n=3,...,N$ and so on. Finally the last user to be decoded, user $N$, experiences no interference. Following that, the data size $Z_n$ of the $\boldsymbol{w}_n$ model parameter, that the $n$-th user transmits within $t$ time duration, should satisfy the following condition
\begin{equation}
Z_n \leq tB\log_2\left(1 + \frac{\frac{E_n}{t}g_n}{\sum_{\{i \in \mathcal{N}|\pi_i>\pi_n\}}\frac{E_i}{t}g_i+BN_0}\right),  \forall n \in \mathcal{N},
\end{equation}
where $B$ is the available bandwidth, $N_0$ denotes the power spectral density, while $E_n$ is the $n$-th user's consumed energy for the aforementioned transmission, with $\frac{E_n}{t}$ being the corresponding  transmit power. Moreover, $g_n=\vert h_n \vert^2d_n^{-2}$ denotes the channel gain, where the complex random variable $h_n \sim \mathcal{CN}(0,1)$
is the small scale fading and $d_n$ is the distance between user
$n$ and the BS. Finally, we assume that the channel remains constant during a communication round and can be perfectly estimated by the BS.
\subsubsection{NOMA with TS} The basic principle of this scheme is that the users' decoding order can change throughout the $t$ transmission time duration, unlike to the NOMA with FDO where a fixed decoding order is considered. According to \cite{goldsmith}, the capacity region which can be achieved by uplink NOMA with TS, known as multiple access channel (MAC) capacity region, is described as \cite{goldsmith}
 \begin{equation}
\sum_{i \in \mathcal{S}}^{}Z_{i} \leq tB\log_2\left(1 + \frac{\sum_{i \in \mathcal{S}}^{}E_{i}g_i}{tBN_0}\right), \quad \forall \mathcal{S} \subseteq \mathcal{N}, \, \mathcal{S} \neq {\O},
\label{MACcr}
\end{equation}
where $\mathcal{S}$ is a non-empty subset of $\mathcal{N}$. Furthermore, it deserves to be mentioned that \eqref{MACcr} also determines the performance of  rate splitting multiple access (RSMA), which is an alternative scheme to achieve any point of the MAC capacity region \cite{rimoldi1996,RSMA}.
\section{Delay Minimization  of a WFL Round}
The aim of the considered optimization problem is to minimize the total delay of a WFL round, i.e., $T=\tau + t$, which is the sum of the computation and transmission latency.  Note that, since the transmit power of the BS is much higher than that of
the users’ and the BS transmits the same message to all users, we ignore the delay of the server for broadcasting
the global parameter. Hereinafter, we focus on a single arbitrary round, while the subsequent analysis can be similarly carried out for each round. Finally, we assume that each user transmits the same data size $Z_n=Z, \, \forall n \in \mathcal{N}$, related with the model parameters, while the considered data size can be predetermined by the central server.
\par
The maximum available CPU clock speed of each user is denoted as $f^{\mathrm{max}}_n$. Thus, it should hold $f_n \leq f^{\mathrm{max}}_n, \,  \forall n \in \mathcal{N}$, which is equivalent to $\tau \geq \underset{n \in \mathcal{N}}{\mathrm{max}}\left(\frac{c_nD_n}{f^{\mathrm{max}}_n}\right)\triangleq a_1$. Moreover, the maximum available energy of each user is $E^{\mathrm{max}}_n$. Hence, it should hold
\begin{equation}
	E^{\mathrm{comp}}_n + E_n=\zeta \frac{c^3_nD^3_n}{\tau^2}+E_n \leq E^{\mathrm{max}}_n, \quad \forall n \in \mathcal{N},
\end{equation}
since the total consumed energy  for both computation and communication purposes, cannot exceed the maximum available energy.
\subsection{CT-NOMA with time-sharing}
The optimization problem for minimizing the total latency in the case of CT-NOMA with TS, can be written as
\begin{equation}
\begin{aligned}
& &  &\underset{\text{\emph{$\tau,t$},{\boldmath{$E$}}}}{\textbf{min}} \quad
\text{\emph{{$\tau+t$}}} \\ 
& & &\textbf{\emph{s.t.}}
\quad \mathrm{C}_{1}: 	\zeta \frac{c^3_nD^3_n}{\tau^2}+E_n \leq E^{\mathrm{max}}_n, \quad \forall n \in \mathcal{N}, \\
& & &\qquad \mathrm{C}_{2}: 	\sum_{i \in \mathcal{S}}^{}Z \leq tB\log_2\left(1 + \frac{\sum_{i \in \mathcal{S}}^{}E_ig_i}{tBN_0}\right), \quad \forall \mathcal{S} \subseteq \mathcal{N}, \\
& & & \qquad   \mathrm{C}_3: \tau \geq a_1, \,t\geq 0,E_n \geq 0, \, \forall n \in \mathcal{N},
\end{aligned}\label{primary}
\end{equation}
where  $\mathrm{C}_{2}$ is related to the users' achievable capacity region in \eqref{MACcr}. It is easy to verify that the problem in \eqref{primary} is jointly convex with respect to (w.r.t.) $\tau,t,\boldsymbol{E}$ and thus can optimally be solved with standard convex-optimization solving methods, such as the interior-point. However, the computational cost to directly solve this by standard methods remains intractable, due to the large number of constraints in $\mathrm{C}_{2}$. More specifically, the number of constraints in $\mathrm{C}_{2}$ is equal to $2^N-1$, since this is the total number of all non-empty subsets $\mathcal{S}$ of the set $\mathcal{N}$. As a result, solving \eqref{primary} for large $N$ comes at the expense of high complexity. To alleviate this burden, an efficient method for solving \eqref{primary} will be proposed. Firstly, we consider a fixed value of $\tau$. Since $\log(\cdot)$ is an ascending  function w.r.t $E_n$, an optimal solution occurs when the constraints $\mathrm{C}_{1}$ are satisfied with equality, i.e., users utilize their whole available energy, which leads to 
\begin{equation}\label{energy_eq}
E_n=E^{\mathrm{max}}_n-\zeta \frac{c^3_nD^3_n}{\tau^2}, \quad  \forall n \in \mathcal{N}. 
\end{equation}
Furthermore, since $E_n \geq 0, \,  \forall n \in \mathcal{N}$, by manipulating \eqref{energy_eq}, yields $\tau \geq \underset{n \in \mathcal{N}}{\mathrm{max}}\left(\sqrt{\zeta\frac{c^3_nD^3_n}{E^{\mathrm{max}}_n}}\right) \triangleq a_2$. Thus, by also recalling that $\tau \geq a_1$, it should finally hold for $\tau$
\begin{equation}
\tau \geq \mathrm{max}\{a_1,a_2\}\triangleq \tau_{\mathrm{low}}.
\end{equation}
Next, after substituting \eqref{energy_eq} and considering a known $\tau \geq \tau_{\mathrm{low}}$, the optimization problem in \eqref{primary} can be reformulated as
\begin{equation}
\begin{aligned}
& &  &\underset{\text{\emph{$t$}}}{\textbf{min}} \quad
\text{\emph{{$t$}}} \\ 
& & &\textbf{\emph{s.t.}} \quad	\sum_{i \in \mathcal{S}}^{}Z \leq tB\log_2\left(1 + \frac{\sum_{i \in \mathcal{S}}^{}A_i}{tBN_0}\right), \quad \forall \mathcal{S} \subseteq \mathcal{N}, \\
\end{aligned}\label{t_primary_problem}
\end{equation}
where $A_i$ is given by
\begin{equation}
	A_i=E_ig_i=\left(E^{\mathrm{max}}_i-\zeta \frac{c^3_iD^3_i}{\tau^2}\right)g_i.
\end{equation}
Note that $A_i$, is constant for a given $\tau$.
Following that, without loss of generality we assume that $A_1\geq A_2\geq...\geq A_N$. According to \cite{diamantoulakis}, when NOMA is combined with the TS scheme and by considering the aforementioned order of $A_i, \, \forall i \in \mathcal{N}$, the achievable capacity region of the users can be equivalently described as 
\begin{equation}
\sum_{i=n}^{N}Z \leq tB\log_2\left(1 + \frac{\sum_{i=n}^{N}A_i}{tBN_0}\right), \quad \forall n \in \mathcal{N},
\label{cr1}
\end{equation}
while by considering that $\sum_{i=n}^{N}Z=Z(N+1-n)$, \eqref{cr1} can be rewritten as
\begin{equation}
	Z \leq \frac{tB\log_2\left(1 + \frac{\sum_{i=n}^{N}A_i}{tBN_0}\right)}{(N+1-n)}, \quad \forall n \in \mathcal{N}.
	\label{cr2}
\end{equation}
It should be noted that the capacity region is now described by $N$ constraints, i.e., $Z$ is bounded by a set of $N$ inequalities.
Next, by exploiting the alternative, but still equivalent representation of the capacity region in \eqref{cr2}, the optimization problem in \eqref{t_primary_problem} can be rewritten as 
\begin{equation}
\begin{aligned}
& &  &\underset{\text{\emph{$t$}}}{\textbf{min}} \quad
\text{\emph{{$t$}}} \\ 
& & & \textbf{\emph{s.t.}} \quad	Z(N+1-n) \leq tB\log_2\left(1 + \frac{\sum_{i=n}^{N}A_i}{tBN_0}\right),
 \forall n \in \mathcal{N}.\\
\end{aligned}\label{t_problem}
\end{equation}
\begin{lemma}
	The optimal value of $t$ for the problem in \eqref{t_problem} can be written as
	\begin{equation}\label{t_solution}
	t^*=\underset{n \in \mathcal{N}}{\mathrm{max}}\left[-\frac{z_n\ln(2)\sum_{i=n}^{N}A_i}{B\left(z_nN_0\ln(2)+\mathcal{W}_{-1}(b_n)\sum_{i=n}^{N}A_i\right)}\right],
	\end{equation}
	where $z_n=(N+1-n)Z$ and $b_n$ is given by
	\begin{equation}
	b_n=-\frac{2^{-\frac{z_nN_0}{\sum_{i=n}^{N}A_i}}z_nN_0\ln(2)}{\sum_{i=n}^{N}A_i},
	\end{equation}
	while $\mathcal{W}_{-1}(\cdot)$ denotes the secondary branch of the Lambert W function.
\end{lemma}
\begin{IEEEproof}
	 We will show that the optimal $t$ is given by the most stringent inequality, among the set of $N$ inequalities in \eqref{t_problem}, when this hold with equality. Firstly, it is straightforward to show that the function 
	\begin{equation}
		f(t)=tB\log_2\left(1 + \frac{\sum_{i=n}^{N}A_i}{tBN_0}\right), \quad \forall t>0, 
	\end{equation}
	is monotonically increasing with respect to $t$. Following that, we assume that $t'$ is optimal and satisfies
	\begin{equation}
		Z(N+1-n) < t'B\log_2\left(1 + \frac{\sum_{i=n}^{N}A_i}{t'BN_0}\right), \, \forall n \in \mathcal{N},
	\end{equation}
	i.e., all constraints are satisfied with strict inequality.
	It will be shown by contradiction that this is not an optimal solution for $t$.
	Since $f(t)$ is increasing w.r.t $t$, there exist an $t^*<t'$, for which at least one inequality constraint from \eqref{t_problem} is satisfied with equality. This observation contradicts the fact that $t'$ is optimal, since the objective is to minimize $t$. Therefore, $t^*$ is given by the most stringent inequality, among the set of $N$ inequalities in \eqref{t_problem}, when this hold with equality.
	After some mathematical manipulations, $t^*$ can be written as in \eqref{t_solution} and the proof is completed.
\end{IEEEproof}
As a matter of fact, the problem can be efficiently solved with closed form solutions for a fixed $\tau$. In the continue, we will show that the global optimal solution can be obtained via the bisection method, in order to find the optimal value of $\tau$. Firstly, it will be shown that $\tau^*$ is bounded, according to the following lemma.
\begin{lemma}
	The optimal $\tau$ is bounded in the following interval
	\begin{equation}
		\tau_{\mathrm{low}} \leq \tau^* \leq \tau_{\mathrm{up}},
	\end{equation}
where $\tau_{\mathrm{up}}=T(\bar{\tau})=\bar{\tau}+t^*(\bar{\tau})$ is the total delay of the communication round that corresponds to an arbitrary feasible value of $\tau$, denoted by $\bar{\tau}$, while  $t^*(\bar{\tau})$ denotes the optimal $t$ for a fixed $\tau = \bar{\tau}$ and is given by \eqref{t_solution}.
\end{lemma}
\begin{IEEEproof}
 As mentioned previously, the optimization problem is feasible when the condition $\tau \geq \tau_{\mathrm{low}}$ is satisfied.  Since $\bar{\tau}$ could be any feasible solution but not necessarily the optimal one, it holds $T(\bar{\tau}) \geq T(\tau^*)$. Following this, by considering that the dedicated time for computations cannot exceed the total delay for any feasible $\tau$, i.e., $\tau < T(\tau)$, we conclude to $\tau^* < T(\tau^*) \leq T(\bar{\tau})$. Thus, the upper bound of $\tau^*$ can be expressed as $\tau_{\mathrm{up}}=T(\bar{\tau})$ and the proof is completed. 
\end{IEEEproof}
\par
 As a result, we have derived the upper and lower bound of $\tau^*$ and now we are ready to apply the bisection method in the considered interval and propose Algorithm 1 for obtaining the optimal solutions that minimize the delay of a communication round. Next, the main steps of the proposed algorithm are discussed. After the necessary initializations in line 1, the bisection method is applied throughout lines 2-13. More specifically, in lines 3- 6, the total delay of the communication round is calculated at the points $\tau=\tau_{\mathrm{m}}$ and $\tau=\tau_{\mathrm{up}}$.  Following that, in lines 7-12, by comparing $T(\tau_{\mathrm{m}})$ and $T(\tau_{\mathrm{up}})$, the bounds of $\tau$ are properly adjusted in each iteration, until the convergence is achieved. It should be highlighted that the joint convexity of the primary problem in \eqref{primary}, w.r.t. all the considered variables, guarantees the convergence of the bisection method to the optimal solution. After the resolution of the optimization problem and the calculation of $\tau^*,t^*$, the optimal $E_n,f_n$ can be given as $E^*_n=E^{\mathrm{max}}_n-\zeta \frac{c^3_nD^3_n}{\tau^{*2}}$ and $f^*_n=\frac{c_nD_n}{\tau^*}$. To this end, the major complexity of Algorithm 1 lies in applying the bisection method and in searching among $N$ values, in order to derive $t^*$ from \eqref{t_solution}. As a result, the complexity can be expressed as $\mathcal{O}\left(N\log_2\left(\frac{\tau_{\mathrm{up}}-\tau_{\mathrm{low}}}{\epsilon}\right)\right)$, where $\epsilon$ denotes the algorithm's tolerance.
\makeatletter
\def\BState{\State\hskip-\ALG@thistlm}
\makeatother
\begin{algorithm}
	\caption{Delay Minimization for CT-NOMA with TS}
	\begin{algorithmic}[1]
		
		\BState  \textbf{Initialize} \text{$\tau_{\mathrm{low}}=\mathrm{max}\{a_1,a_2\},\,\tau_\mathrm{m}=\bar{\tau},\,\tau_{\mathrm{up}}=T(\bar{\tau}),\,\epsilon;$}
		\BState \textbf{while} \text{$\tau_{\mathrm{up}}-\tau_{\mathrm{low}}>\epsilon$} \textbf{do}
		\BState \quad \textbf{Set} $\tau=\tau_\mathrm{m},$ \text{ and derive} $t^*(\tau_\mathrm{m})$ \text{from \eqref{t_solution}}
		\BState \quad \textbf{Set} $T(\tau_\mathrm{m})=\tau_\mathrm{m}+t^*(\tau_\mathrm{m});$ 
		\BState \quad \textbf{Set} $\tau=\tau_\mathrm{up},$ \text{and derive} $t^*(\tau_\mathrm{up})$ \text{from \eqref{t_solution}}
		\BState \quad \textbf{Set} $T(\tau_\mathrm{up})=\tau_\mathrm{up}+t^*(\tau_\mathrm{up});$
		\BState \quad \textbf{if} $T(\tau_\mathrm{m})<T(\tau_\mathrm{up})$
		\BState \qquad $\tau_\mathrm{up}=\tau_\mathrm{m};$
		\BState \quad \textbf{else}
		\BState \qquad $\tau_\mathrm{low}=\tau_\mathrm{m};$
		\BState \quad \textbf{end if}
		\BState \quad $\tau_\mathrm{m}=\frac{\tau_\mathrm{low}+\tau_\mathrm{up}}{2};$
		\BState \textbf{end while}
		\BState \textbf{Output} $\tau^* =\tau_\mathrm{up}, \, t^*=t^*(\tau_\mathrm{up}),\, T^*=\tau^*+t^*;$
	\end{algorithmic}\label{alg}
\end{algorithm}

\subsection{CT-NOMA with fixed decoding order}
The optimization problem for minimizing the total delay in the case of CT-NOMA with FDO, can be written as
\begin{equation}
\begin{aligned}
& &  &\underset{\text{\emph{$\tau,t$},{\boldmath{$E,\pi$}}}}{\textbf{min}} \quad
\text{\emph{{$\tau+t$}}} \\ 
& & &\textbf{\emph{s.t.}}
\quad \mathrm{C}_{1}: 	\zeta \frac{c^3_nD^3_n}{\tau^2}+E_n \leq E^{\mathrm{max}}_n, \quad \forall n \in \mathcal{N}, \\
& & &\qquad \mathrm{C}_{2}: 	Z \leq tB\log_2\left(1 + \frac{\frac{E_n}{t}g_n}{\sum_{\{i \in \mathcal{N}|\pi_i>\pi_n\}}\frac{E_i}{t}g_i+BN_0}\right)\\
 & & &\qquad \qquad \forall n \in \mathcal{N}, \\
& & & \qquad   \mathrm{C}_3: \tau \geq a_1, \,t\geq 0,E_n \geq 0, \, \forall n \in \mathcal{N}, \, \boldsymbol{\pi}\in \Pi,
\end{aligned}\label{fixed_decoding}
\end{equation}
where the permutation $\boldsymbol{\pi}$ belongs to the set $\Pi$, defined as the set of all possible decoding orders among the $N$ users.
 Note that the total number of permutations, i.e., all possible decoding orders, is equal to $\vert\Pi\vert=N!$. Therefore, the optimization problem in \eqref{fixed_decoding} is of a combinatorial nature. However, an exhaustive search among all possible decoding orders is prohibitive, due to high complexity. To this end, we adopt an ascending decoding order w.r.t. to the channel gains, according to which the messages of the users with  weaker channel gains are exposed to less interference during the decoding process. It is notable that this a common selection for uplink NOMA systems, since it provides fairness among users without reducing the sum rate \cite{yang2016,diamantoulakis}. Accordingly, by considering  without loss of generality that $g_1\geq g_2\geq...\geq g_N$, we set $\pi_1<\pi_2<...<\pi_N$. As a result, $\mathrm{C}_{2}$ of \eqref{fixed_decoding} can be reformulated as
\begin{equation}
	Z \leq tB\log_2\left(1 + \frac{E_ng_n}{\sum_{i=n+1}^{N}E_ig_i+tBN_0}\right), \quad \forall n \in \mathcal{N}.
	\label{C2initial}
\end{equation}
By substituting \eqref{C2initial} in \eqref{fixed_decoding}, the later can be written as
\begin{equation}
\begin{aligned}
& &  &\underset{\text{\emph{$\tau,t$},{\boldmath{$E$}}}}{\textbf{min}} \quad
\text{\emph{{$\tau+t$}}} \\ 
& & &\textbf{\emph{s.t.}}
\quad \mathrm{C}_{1}: 	\zeta \frac{c^3_nD^3_n}{\tau^2}+E_n \leq E^{\mathrm{max}}_n, \quad \forall n \in \mathcal{N}, \\
& & &\qquad \mathrm{C}_{2}: Z \leq tB\log_2\left(1 + \frac{E_ng_n}{\sum_{i=n+1}^{N}E_ig_i+tBN_0}\right),\\
& & &\qquad \qquad \forall n \in \mathcal{N}, \\
& & & \qquad   \mathrm{C}_3: \tau \geq a_1, \,t\geq 0,E_n \geq 0, \, \forall n \in \mathcal{N},
\end{aligned}\label{fixed_decoding2}
\end{equation}
which is non-convex due to the coupling of $t$ and $\boldsymbol{E}$ in constraint $\mathrm{C}_{2}$. In order to address the non-convexity issues, we first rewrite $\mathrm{C}_{2}$ as
\begin{equation}
	2^{\frac{Z}{tB}} \leq \frac{E_ng_n}{\sum_{i=n+1}^{N}E_ig_i+tBN_0}, \, \forall n \in \mathcal{N}.
\end{equation}
In the continue we set $E_n\triangleq \exp(\tilde{E}_n),\, \forall n \in \mathcal{N}$ and $t\triangleq \exp(\tilde{t})$. After some mathematical manipulations, $\mathrm{C}_{2}$ can be written as
\begin{equation}
\begin{aligned}
& & &	\ln\left(2^{\frac{Z}{B}\exp(-\tilde{t})}-1\right)\\
& & &+\ln\left(\sum_{i=n+1}^{N}\exp(\tilde{E}_i-\tilde{E}_n)g_i+\exp(\tilde{t}-\tilde{E}_n)BN_0\right)\\
& & & \leq \ln(g_n), \quad \forall n \in \mathcal{N}.
\end{aligned}
\end{equation}
By exploiting the aforementioned formulation of $\mathrm{C}_{2}$, the optimization problem in \eqref{fixed_decoding2} can be transformed to the following equivalent one:
\begin{equation}
\begin{aligned}
& &  &\underset{\text{\emph{$\tau,\tilde{t}$},{\boldmath{$\tilde{E}$}}}}{\textbf{min}} \quad
\text{\emph{{$\tau+\exp(\tilde{t})$}}} \\ 
& & &\textbf{\emph{s.t}}
\quad \mathrm{C}_{1}: 	\zeta \frac{c^3_nD^3_n}{\tau^2}+\exp(\tilde{E}_n) \leq E^{\mathrm{max}}_n, \quad \forall n \in \mathcal{N}, \\
& & &\qquad \mathrm{C}_{2}: 	\ln\left(2^{\frac{Z}{B}\exp(-\tilde{t})}-1\right)
+\ln\Biggl(\sum_{i=n+1}^{N}\exp(\tilde{E}_i-\tilde{E}_n)g_i\\
& & & \qquad \qquad +\exp(\tilde{t}-\tilde{E}_n)BN_0\Biggr) -\ln(g_n) \leq 0, \, \forall n \in \mathcal{N}, \\
& & & \qquad   \mathrm{C}_3: \tau \geq a_1, \,\tilde{t}\in \mathbb{R},{\tilde{E}_n} \in \mathbb{R}.
\end{aligned}\label{fixed_decoding3}
\end{equation}
It can be easily shown that the optimization problem in \eqref{fixed_decoding3} is convex, while the proof is omitted for brevity. Thus, it can efficiently be solved with standard convex optimization methods.
\section{Performance Evaluation and Discussion}
\begin{figure}[h!]
	\centering
	\includegraphics[width=0.8\linewidth]{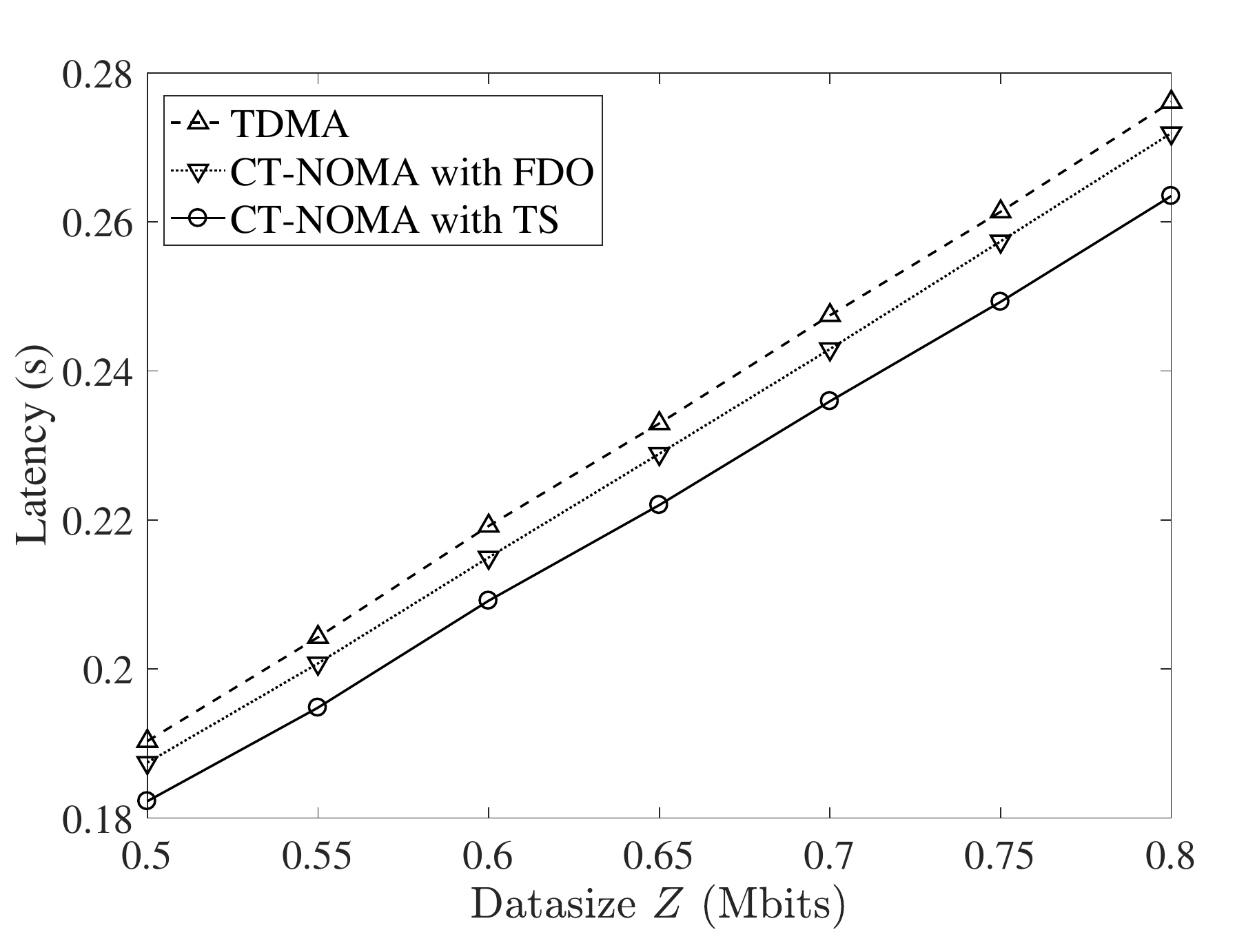}
		\vspace{-0.1in}
	\caption{Impact of the users' parameter data size on latency, with $E^{\mathrm{max}}_n=2\, \mathrm{Joule}$.}
		\vspace{-0.1in}
	\label{datasize}
\end{figure}

Table I summarizes the simulations' parameters setup. The following figures have been extracted by means of Monte Carlo.
\begin{figure}[h!]
	\centering
	\includegraphics[width=0.8\linewidth]{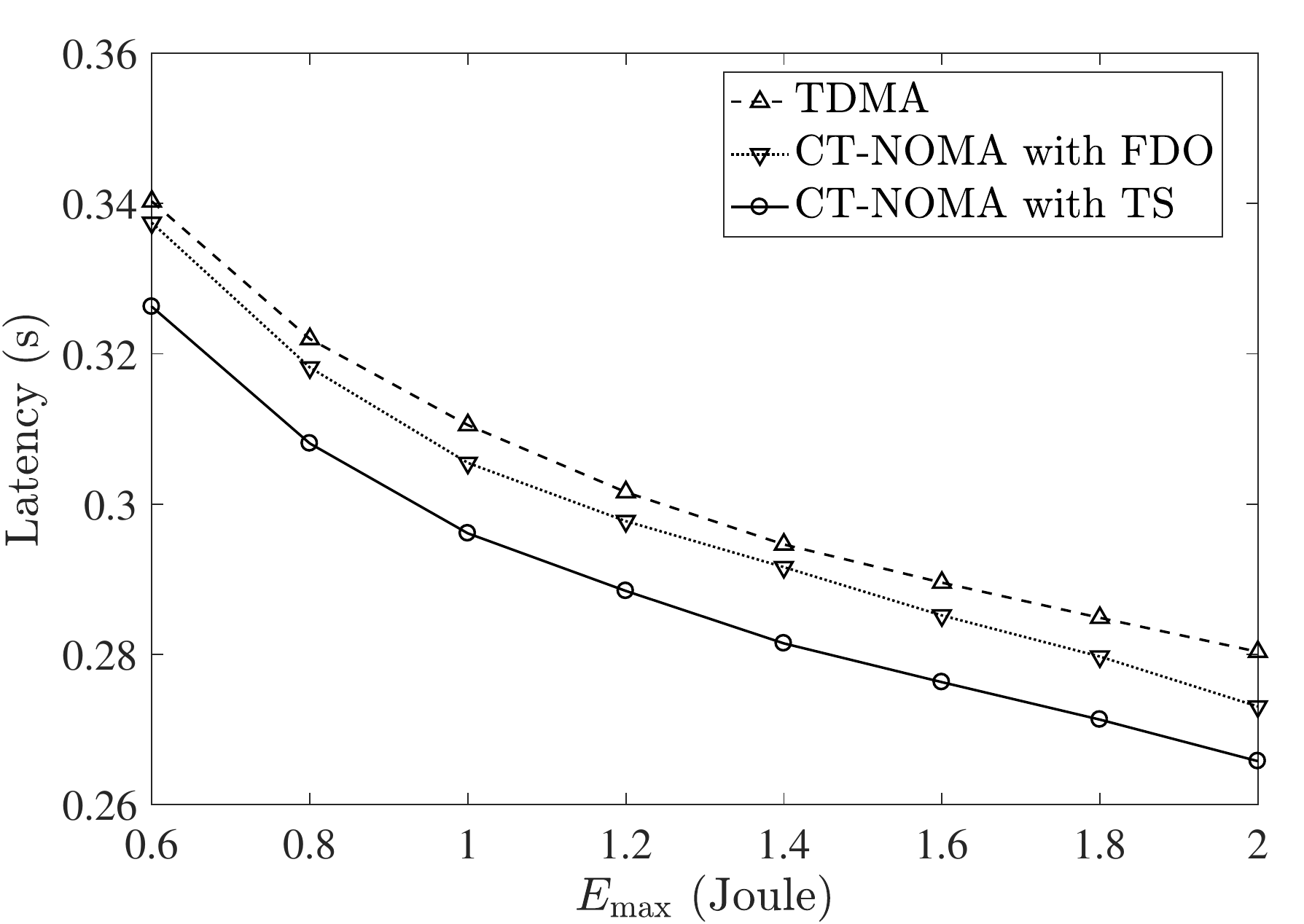}
	\vspace{-0.1in}
	\caption{Impact of the users' maximum available energy on latency, with $Z=0.8\mathrm{Mbits}$.}
	\vspace{-0.1in}
	\label{latency}
\end{figure}
In Fig. \ref{latency}, the impact of the users' maximum available energy on the average latency during a WFL round is depicted. We use the TDMA-based protocol as benchmark  and compare it with the CT-NOMA protocol in terms of delay reduction, taking into account both considered detections schemes for NOMA. It can been observed that CT-NOMA with FDO slightly outperforms TDMA. However, CT-NOMA with TS clearly dominates both TDMA and CT-NOMA with FDO. This is due to the fact that CT-NOMA with TS can achieve any point of the capacity region, in contrast to CT-NOMA with FDO which achieves only the corner points. As a result, the efficient exploitation of the capacity region, as well as subsequently the flexible interplay among users' data rates that CT-NOMA with TS enables, lead to decreased delay during the WFL round. Moreover, in Fig. \ref{datasize}, the impact of the users' parameter data size on latency is illustrated. The superiority of CT-NOMA with TS against CT-NOMA with FDO, and TDMA is again corroborated. Furthermore, CT-NOMA with FDO presents a slightly enhanced performance in comparison with TDMA.
\par
By evaluating the demonstrated results, it is evident that CT-NOMA has the potential to provide decreased latency during a WFL round and subsequently accelerate the training process, which is an important requirement for the efficient integration of WFL in 6G. Moreover, since NOMA can achieve any point of the capacity region through the use of time-sharing (or rate splitting) technique, it can serve in future research as a performance upper bound in order to evaluate other multiple access schemes for WFL. Furthermore, the proposed protocol could serve as a baseline for more sophisticated network implementations, such as hybrid NOMA/OMA configurations, which could possibly enhance the scalability of WFL.
\begin{table}
	\caption{Simulation Settings}
	\centering
	\begin{tabular}{ p{1.5cm}|p{1.5cm}||p{1.5cm}|p{2cm}}
		\hline
		\centering
		\textbf{Parameter} & \textbf{Value} & \textbf{Parameter} & \textbf{Value}\\
		\hline\hline
		$f^{\mathrm{max}}_n$ & $1.5\mathrm{GHz}$ & $D_n$ & 1Mbit \\
		\hline
		$B$ & 1MHz & $N_0$ & 174dBm/Hz \\
		\hline
		$\zeta$ & $10^{-27}$ & $c_n$(cycles) & $\sim \mathcal{U}(10,40)$ \\
		\hline
		$N$ & 10 users & $d_n$ & $\sim \mathcal{U}(0,500\mathrm{m})$ \\
		\hline

	\end{tabular}
	\vspace{-0.2in}
\end{table}
	\vspace{-0.2in}

\bibliographystyle{IEEEtran}
\bibliography{bibl}
\end{document}